\documentclass[aps,prd,showpacs,preprintnumbers,nofootinbib]{revtex4}
\usepackage[dvips]{graphicx}
\usepackage{bm,latexsym,amsmath,amssymb,amsfonts}
\usepackage{color}

\newcommand{\6}{\partial}
\begin{document}

\title{Angular momentum at null infinity in five dimensions}

\author{Kentaro Tanabe}
\email{tanabe@yukawa.kyoto-u.ac.jp}
\affiliation{Yukawa Institute for Theoretical Physics, Kyoto University, Kyoto 606-8502, Japan}
\author{Norihiro Tanahashi}
\email{tanahashi@ms.physics.ucdavis.edu}
\affiliation{Department of Physics, University of California, Davis, CA 95616, USA}
\author{Tetsuya Shiromizu}
\email{shiromizu@tap.scphys.kyoto-u.ac.jp}
\affiliation{Department of Physics, Kyoto University, Kyoto 606-8502, Japan}
\begin{abstract}
In this paper, using the Bondi coordinates, 
we discuss the angular momentum at null infinity in five dimensions 
and address the Poincare covariance of the Bondi mass and angular momentum. 
We also show the angular momentum loss/gain law due to gravitational waves. 
In four dimensions, the angular momentum at null infinity has the supertranslational 
ambiguity and then it is known that we cannot construct well-defined angular momentum 
there. On the other hand, we would stress that 
we can define angular momentum at null infinity without any ambiguity in higher 
dimensions. This is because of the non-existence of supertranslations in higher 
dimensions. 
\end{abstract}
\maketitle

\section{introduction}
Inspired by the recent progress of the string theory, 
the importance of the gravity theory in higher dimensional space-times is steadily growing. 
However, there are still many remaining issues to be investigated in higher dimensions. 
One issue among them is the 
asymptotic structure. For asymptotically flat space-times, the asymptotic 
structure is defined at spatial  and null infinites. The asymptotic 
structure at spatial infinity (spi) is well-defined by conformal embedding 
in four~\cite{Ashtekar:1978, Geroch:1977jn, Geroch:1972} and higher 
dimensions~\cite{Tanabe:2009xb}. The asymptotic structure at null infinity in four 
dimensions is well studied by many authors~\cite{Geroch:1977jn, Bondi:1962px, 
Sachs:1962wk, Sachs:1962, Penrose:1962ij, Newman:1966ub, Penrose:1965am, 
Geroch:1978ub, Geroch:1978ur}.
On the other hand, there 
are only a few work about the asymptotic structure at null infinity in higher dimensions. 
Indeed, asymptotic flatness has been defined 
by using conformal completion method in only even dimensions~\cite{Hollands:2003ie, 
Hollands:2003xp, Ishibashi:2007kb} and by using the Bondi coordinates in five 
dimensions~\cite{Tanabe:2009va}. 

In four dimensions, asymptotic structure at null infinity is often studied 
by using the conformal embedding~\cite{Penrose:1962ij}. 
In this method, 
we introduce the conformal factor $\Omega\sim1/r$ and use $\Omega$ as a
coordinate near null infinity $\Omega=0$. 
In four dimensions, this method provided us successful results for the 
analysis of asymptotic structure. 
However, it turned out that 
this method does not work well in higher dimensions.
In fact, we cannot guarantee the smoothness of gravitational fields 
at null infinity in the coordinate of $\Omega$, 
particularly in odd dimensions~\cite{Hollands:2003ie, Hollands:2003xp, 
Ishibashi:2007kb, Hollands:2004ac}. This is because gravitational wave 
behaves $1/r^{(D-2)/2}$ near null infinity in $D$ dimensions. This means that 
the conformal completion method might not be the best way to 
study the issue.
Then we must solve the Einstein equations directly to study how the 
gravitational field is expanded near null infinity in odd dimensions. 
By introducing the Bondi coordinates instead, we can study the asymptotic structure at null 
infinity even in odd dimensions without assuming the smoothness of the gravitational 
fields. 
Indeed, in this way, we can derive the Bondi mass loss law by gravitational waves 
in five dimensions. We can also show that the regularity of gravitational field at null infinity 
and that asymptotic symmetry is the Poincare group, i.e., there are no supertranslations 
at null infinity in five dimensions~\cite{Tanabe:2009va}
while there are always supertranslations in four dimensions. 

The asymptotic symmetry at null infinity in four dimensions is semi-direct product 
of the Lorentz group and the supertranslational group, which is an infinite dimensional 
translational group. The presence of supertranslations implies the infinite number 
of the direction of translation, while the Poincare group has only four directions in 
four dimensions. Because of this infinite directions of translation, we cannot construct well-defined 
angular momentum in four dimensions. There are many attempts to define of angular momentum at null infinity 
in four dimensions, whereas all those definitions are suffered from 
supertranslational ambiguity~\cite{Tamburino:1966zz, Winicour:1968, Prior:1977, 
Streubel:1978, Winicour:1980, Geroch:1981ut, Dray:1984}. 
On the other hands, in five dimensions, since asymptotic symmetry is 
the Poincare group, we can expect that angular momentum at null infinity 
can be defined without any ambiguities. The purpose of this paper is to discuss 
angular momentum at null infinity in five dimensions. We will see that angular momentum 
can be defined well and show the Poincare covariance of the Bondi mass and angular 
momentum. We also look at the angular momentum loss/gain properties due to gravitational 
waves. 

The rest of this paper is organized as follows. In the next section, we review our previous work 
\cite{Tanabe:2009va}. Therein we introduce the Bondi coordinates and solve the Einstein equations 
near null infinity. In section~\ref{Sec:AQ} we define the Bondi mass and the Bondi angular 
momentum, and derive the Bondi mass loss law and angular momentum loss/gain law by 
gravitational waves. In section~\ref{Sec:AS} we show the Poincare covariance of Bondi mass and 
angular momentum. In section~\ref{Sec:Summary} we summarize our paper 
and discuss the extension to higher dimensions than seven. 
In Appendix A, for comparison, we will consider angular momentum at null infinity 
in four dimensions using the Bondi coordinates. This is 
because we cannot find references which address angular momentum using 
the Bondi coordinate. We will show that there is always the supertranslational 
ambiguity in the angular momentum.

\section{The Bondi coordinates and the Einstein equations}
\label{Sec:Bondi}

In this section we review our previous work of Ref.~\cite{Tanabe:2009va} on
the asymptotic structure at null infinity in five dimensions. Therein we 
used the Bondi coordinates, which will be useful to study the behavior of 
gravitational fields via solving of the Einstein equation.

\subsection{Bondi coordinates}
In the Bondi coordinates $x^{a}=(u,r,\theta,\phi,\psi)$ the metric can be written 
as
%
\begin{equation}
ds^{2}\,=\,-\frac{Ve^{B}}{r^2}du^{2}-2e^{B}dudr+r^{2}h_{AB}(dx^{A}+U^{A}du)(dx^{B}+U^{B}du),
\end{equation} 
%
where\footnote{$x^{A}=(\theta,\phi,\psi)$}
%
\begin{equation}
h_{AB}\,=\,
\begin{pmatrix}
e^{C_{1}} & \sin\theta\sinh D_{1} & \cos\theta\sinh D_{2} \\
\sin\theta\sinh D_{1} & e^{C_{2}}\sin^{2}\theta & \sin\theta\cos\theta \sinh D_{3} \\
\cos\theta\sinh D_{2} & \sin\theta\cos\theta \sinh D_{3} & e^{C_{3}}\cos^{2}\theta
\end{pmatrix},
\end{equation} 
%
and we adopted the gauge condition satisfying $\det h_{AB}\,=\,\sin^{2}\theta 
\cos^{2}\theta$. $u=\text{const.}$ are null hypersurfaces and the 
periods of the coordinates $\theta$, $\phi$ and $\psi$ are $\pi/2$, $2\pi$ and 
$2\pi$, respectively.  From the gauge condition, $e^{C_{3}}$ can be 
written as
%
\begin{equation}
e^{C_{3}}\,=\,\frac{1+e^{C_{2}}\sinh^{2}D_{2}+e^{C_{1}}\sinh^{2}D_{3}-2\sinh D_{1}\sinh D_{2}\sinh D_{3}}{e^{C_{1}+C_{2}}-\sinh^{2}D_{1}}.
\end{equation} 
%
Then $h_{AB}$ have five functional freedom. In the following we will identify  
$C_{1},C_{2},D_{1},D_{2},D_{3}$ as those freedom. In this coordinate system, 
null infinity is represented by $r\,=\,\infty$ and the metric at null infinity is
%
\begin{equation}
ds^{2}\,=\,-du^{2}-2dudr+r^2(d\theta^{2} +\sin^{2}\theta d\phi^{2} +\cos^{2}\theta d\psi^{2}). 
\label{bgmetric}
\end{equation} 
%

\subsection{The Einstein equations}

To investigate the asymptotic structure at null infinity, we have to solve the 
Einstein equations near null infinity. Here note that five dimensional space-times have 
five degree of freedom of gravitational fields. If we identify $h_{AB}$ as the freedom of 
gravitational field, $C_{1},C_{2},D_{1},D_{2},D_{3}$ can be expanded as 
%
\begin{gather}
C_{1}(u,r,x^{A})\,=\,\frac{C_{11}(u,x^{A})}{r\sqrt{r}} +\frac{C_{12}(u,x^{A})}{r^{2}}
           +\frac{C_{13}(u,x^{A})}{r^{2}\sqrt{r}}+\frac{C_{14}(u,x^{A})}{r^{3}}+O(r^{-7/2}) \label{bbc1}\\ 
C_{2}(u,r,x^{A})\,=\,\frac{C_{21}(u,x^{A})}{r\sqrt{r}} +\frac{C_{22}(u,x^{A})}{r^{2}}
           +\frac{C_{23}(u,x^{A})}{r^{2}\sqrt{r}}+\frac{C_{24}(u,x^{A})}{r^{3}}+O(r^{-7/2}) \\ 
D_{1}(u,r,x^{A})\,=\,\frac{D_{11}(u,x^{A})}{r\sqrt{r}} +\frac{D_{12}(u,x^{A})}{r^{2}}
           +\frac{D_{13}(u,x^{A})}{r^{2}\sqrt{r}}+\frac{D_{14}(u,x^{A})}{r^{3}}+O(r^{-7/2}) \\ 
D_{2}(u,r,x^{A})\,=\,\frac{D_{21}(u,x^{A})}{r\sqrt{r}} +\frac{D_{22}(u,x^{A})}{r^{2}}
           +\frac{D_{23}(u,x^{A})}{r^{2}\sqrt{r}}+\frac{D_{24}(u,x^{A})}{r^{3}}+O(r^{-7/2}) \\ 
D_{3}(u,r,x^{A})\,=\,\frac{D_{31}(u,x^{A})}{r\sqrt{r}} +\frac{D_{32}(u,x^{A})}{r^{2}}
           +\frac{D_{33}(u,x^{A})}{r^{2}\sqrt{r}}+\frac{D_{34}(u,x^{A})}{r^{3}}+O(r^{-7/2}) \label{bbc2}.
\end{gather} 
%
The Einstein equations $R_{rr}=0$, $R_{rA}=0$ and the trace part of 
$R_{AB}=0$ determine the behaviors of $B$, $U^{A}$
and $V$ near null infinity as
%
\begin{eqnarray}
\frac{V}{r^2} &=& 1+\frac{V_{1}(u,x^{A})}{r\sqrt{r}} -\frac{m(u,x^{A})}{r^2} +O(r^{-5/2}) \\
B &=& \frac{B_{1}(u,x^{A})}{r^3} +O(r^{-4}) \\
U^{A} &=& \frac{U_{1}^{A}(u,x^{A})}{r^2\sqrt{r}} + \frac{U_{2}^{A}(u,x^{A})}{r^3} 
+ \frac{U_{3}^{A}(u,x^{A})}{r^3\sqrt{r}}+\frac{U_{4}^{A}(u,x^{A})}{r^4} +O(r^{-9/2}) \\
h_{AB} &=& h^{(0)}_{AB} + \frac{1}{r\sqrt{r}}h^{(1)}_{AB} + \frac{1}{r^2}h^{(2)}_{AB} 
+ \frac{1}{r^2\sqrt{r}}h^{(3)}_{AB}+O(r^{-5/2}) ,
\end{eqnarray} 
%
where
%
\begin{equation}
h^{(0)}_{AB}\,=\,
\begin{pmatrix}
1 & 0 & 0 \\
0 & \sin^{2}\theta & 0 \\
0 & 0 & \cos^{2}\theta
\end{pmatrix}
\end{equation} 
%
and 
%
\begin{equation}
h^{(n)}_{AB}\,=\,
\begin{pmatrix}
C_{1n}(u,x^{A}) & \sin\theta D_{1n}(u,x^{A}) & \cos\theta D_{2n}(u,x^{A}) \\
\sin\theta D_{1n}(u,x^{A}) & C_{2n}(u,x^{A})\sin^{2}\theta & \sin\theta
\cos\theta  D_{3n}(u,x^{A}) \\
\cos\theta D_{2n}(u,x^{A}) & \sin\theta\cos\theta  D_{3n}(u,x^{A}) & 
-(C_{1n}(u,x^{A})+C_{2n}(u,x^{A}))\cos^{2}\theta
\end{pmatrix}
\end{equation} 
%
for $n=1,2,3$. The coefficients in these expansions are all written by 
the gravitational fields $C_{1},C_{2},D_{1},D_{2},D_{3}$. From $R_{rr}\,=\,0$, 
we have 
%
\begin{equation}
B_{1}(u,x^{A})\,=\,
-\frac{1}{8}(C_{11}^2+C_{11}C_{21}+C_{21}^2+D_{11}^2+D_{21}^2+D_{31}^2).
\label{B1}
\end{equation} 
%
From $R_{rA}=0$, 
%
\begin{eqnarray}
U_{1}^{\theta} &=& \frac{2}{5}\left[ \frac{1}{\sin\theta\cos^{2}\theta}\frac{\partial}{\partial\theta}
(\sin\theta\cos^{2}\theta C_{11})+\frac{1}{\sin\theta}\frac{\partial}{\partial\phi} D_{11} 
+\frac{1}{\cos\theta}\frac{\partial}{\partial\psi}D_{21}-\frac{1}{\sin\theta\cos\theta}C_{21} \right] 
\label{U11}\\
\sin^{2}\theta U_{1}^{\phi} &=& \frac{2}{5}\left[ \frac{1}{\sin\theta\cos\theta}\frac{\partial}
{\partial\theta}(\sin^{2}\theta\cos\theta D_{11})+\frac{\partial}{\partial\phi}C_{21} 
+\tan\theta\frac{\partial}{\partial\psi}D_{31}  \right] 
\label{U21}\\
\cos^{2}\theta U_{1}^{\psi} &=& \frac{2}{5}\left[ \frac{1}{\sin\theta\cos\theta}\frac{\partial}
{\partial\theta}(\sin\theta\cos^{2}\theta D_{21})+\cot\theta\frac{\partial}{\partial\phi}D_{31} 
-\frac{\partial}{\partial\psi}(C_{11}+C_{21})  \right] ,
\label{U31}
\end{eqnarray} 
%
%
\begin{eqnarray}
U_{2}^{\theta} &=& \frac{2}{3}\left[ \frac{1}{\sin\theta\cos^{2}\theta}\frac{\partial}{\partial\theta}
(\sin\theta\cos^{2}\theta C_{12})+\frac{1}{\sin\theta}\frac{\partial}{\partial\phi} D_{12} 
+\frac{1}{\cos\theta}\frac{\partial}{\partial\psi}D_{22}-\frac{1}{\sin\theta\cos\theta}C_{22} \right] 
\label{U12}\\
\sin^{2}\theta U_{2}^{\phi} &=& \frac{2}{3}\left[ \frac{1}{\sin\theta\cos\theta}\frac{\partial}
{\partial\theta}(\sin^{2}\theta\cos\theta D_{12})+\frac{\partial}{\partial\phi}C_{22} +\tan\theta
\frac{\partial}{\partial\psi}D_{32}  \right] 
\label{U22}\\
\cos^{2}\theta U_{2}^{\psi} &=& \frac{2}{3}\left[ \frac{1}{\sin\theta\cos\theta}\frac{\partial}
{\partial\theta}(\sin\theta\cos^{2}\theta D_{22})+\cot\theta\frac{\partial}{\partial\phi}D_{32} 
-\frac{\partial}{\partial\psi}(C_{12}+C_{22})  \right] 
\label{U32},
\end{eqnarray} 
%
and
%
\begin{eqnarray}
U_{3}^{\theta} &=& \frac{10}{7}\left[ \frac{1}{\sin\theta\cos^{2}\theta}\frac{\partial}{\partial\theta}
(\sin\theta\cos^{2}\theta C_{13})+\frac{1}{\sin\theta}\frac{\partial}{\partial\phi }D_{13} 
+\frac{1}{\cos\theta}\frac{\partial}{\partial\psi}D_{23}-\frac{1}{\sin\theta\cos\theta}C_{23} \right] 
\label{U13}\\
\sin^{2}\theta U_{3}^{\phi} &=& \frac{10}{7}\left[ \frac{1}{\sin\theta\cos\theta}\frac{\partial}
{\partial\theta}(\sin^{2}\theta\cos\theta D_{13})+\frac{\partial}{\partial\phi}C_{23} +\tan\theta
\frac{\partial}{\partial\psi}D_{33}  \right] 
\label{U23}\\
\cos^{2}\theta U_{3}^{\psi} &=& \frac{10}{7}\left[ \frac{1}{\sin\theta\cos\theta}\frac{\partial}
{\partial\theta}(\sin\theta\cos^{2}\theta D_{23})+\cot\theta\frac{\partial}{\partial\phi}D_{33} 
-\frac{\partial}{\partial\psi}(C_{13}+C_{23})  \right] 
\label{U33}.
\end{eqnarray} 
%
From the trace part of $R_{AB}=0$, we can obtain
%
\begin{equation}
V_{1}(u,x^{A})\,=\,-\frac{2}{3}\left( \frac{1}{\sin\theta\cos\theta}\frac{\partial}{\partial\theta}
(\sin\theta\cos\theta U_{1}^{\theta})+\frac{\partial}{\partial\phi}U_{1}^{\phi} +\frac{\partial}
{\partial\psi}U_{1}^{\psi} \right) \label{vsol}. 
\end{equation} 
%
These solutions will be needed when we define the Bondi mass and angular momentum 
in the next section and confirm the Poincare covariance of the Bondi mass and angular momenta 
in section~\ref{Sec:AS}. 

The equations in the traceless part of $R_{AB}=0$ describe the evolution 
of $h_{AB}^{(n)}~(n>1)$ along the $u$-direction. 
Since $\partial h^{(1)}_{AB}/\partial u$ does not appear in those equations, 
we may set it arbitrarily on each $u=\text{const.}$ hypersurfaces. 
This degree of freedom can be regarded as the degree of freedom of gravitational waves.
Furthermore, we can see that $h^{(2)}_{AB}$ are time independent 
$\partial h^{(2)}_{AB}/\partial u=0$ \cite{Tanabe:2009va}. 
This fact will play a key role when showing the Poincare covariance of the Bondi momentum. 
The functions $m(u,x^{A})$ and $U^{A}_{4}(u,x^{A})$ appear as
the integration constants of $r$-integration of the equations for each $(u,x^{A})$, 
that is, they are free 
functions on $u=\text{const.}$ hypersurfaces. As seen later, these functions represent 
the energy and angular momenta contained in $u=\text{const.}$ hypersurfaces.

\section{Asymptotic quantity}
\label{Sec:AQ}

In this section, we define the Bondi mass and angular momenta. 
The normalization factors are determined so that these quantities coincide to Arnowitt-Deser-Misner (ADM)
quantities at spatial infinity~\cite{Harmark:2004rm}.

\subsection{The Bondi mass and the Bondi momentum}

We firstly define the Bondi mass and momentum from the asymptotic behaviors of 
metric components. Since $g_{uu}$ is expanded near null infinity as 
%
\begin{equation}
g_{uu}\,=\,-1-\frac{V_{1}(u,x^{A})}{r\sqrt{r}}+\frac{m(u,x^{A})}{r^{2}}+O(r^{-5/2}), 
\label{guu}
\end{equation} 
%
it is natural to define Bondi mass and momentum as 
%
\begin{equation}
M_{\text{Bondi}}(u)\,=\,\frac{3}{16\pi}\int_{S^{3}}m(u,x^{A})d\Omega
\end{equation} 
%
and
%
\begin{equation}
P^{i}_{\text{Bondi}}(u)\,=\,\frac{3}{16\pi}\int_{S^{3}}m(u,x^{A})\hat{x}^{i}d\Omega,
\end{equation} 
%
respectively. In the aboves 
$\hat{x}^{i}=(\hat{x},\hat{y},\hat{z},\hat{w})=
(\sin\theta\cos\phi, \sin\theta\sin\phi, \cos\theta\cos\psi, 
\cos\theta\sin\psi)$, which are $l=1$ modes of the scalar harmonics on $S^{3}$ and 
$d\Omega=\sin\theta\cos\theta d\theta d\phi d\psi$. The 
Bondi five-momentum $P^{a}_{\text{Bondi}}$ are defined by 
$P^{a}_{\text{Bondi}}=(M_{\text{Bondi}},P^{i}_{\text{Bondi}})$.

From the Einstein equation of $R_{uu}=0$, we can obtain the Bondi mass loss law such as
%
\begin{eqnarray}
\frac{d}{du}M_{\text{Bondi}}&=&
\frac{3}{16\pi}\int_{S^{3}}\frac{\partial m(u,x^{A})}{\partial u}d\Omega \notag\\
&=&
-\frac{1}{16\pi}\int_{S^{3}}\left\{ 
\left(\frac{\partial C_{11}}{\partial u}\right)^2 +\frac{\partial C_{11}}
{\partial u}\frac{\partial C_{21}}{\partial u} +\left(\frac{\partial C_{21}}{\partial u}\right)^2 
+\left(\frac{\partial D_{11}}{\partial u}\right)^2 
+\left(\frac{\partial D_{21}}{\partial u}\right)^2 +\left(\frac{\partial D_{31}}
{\partial u}\right)^2\right. \notag\\
&&~~~~~~~~~~~~~~~~~~~~~~~~~~~~~~~~~~~~~~~~~~~~~~~\left. -
\frac{2}{\sin\theta\cos\theta}\frac{\partial}{\partial\theta}\left(\sin\theta\cos\theta \frac{\partial 
U^{\theta}_{2}}{\partial u}\right) -2\frac{\partial^{2}}{\partial\phi\partial u}U^{\phi}_{2}
-2\frac{\partial^{2}}{\partial\psi\partial u}U^{\psi}_{2}
\right\} d\Omega \notag \\\
&=&
-\frac{1}{16\pi}\int_{S^{3}}\left\{ 
\left(\frac{\partial C_{11}}{\partial u}\right)^2 +\frac{\partial C_{11}}
{\partial u}\frac{\partial C_{21}}{\partial u} +\left(\frac{\partial C_{21}}{\partial u}\right)^2 
+\left(\frac{\partial D_{11}}{\partial u}\right)^2 
+\left(\frac{\partial D_{21}}{\partial u}\right)^2 +\left(\frac{\partial D_{31}}
{\partial u}\right)^2 \right\}d\Omega \notag\\
& \leq & 0. \label{masslosslaw}
\end{eqnarray} 
%
Thus, it turns out that the Bondi mass always decreases due to 
gravitational waves. The total derivative terms in
this integral have no contributions to the Bondi mass loss. 

We comment on the finiteness of the Bondi mass. In the conformal completion method, we usually define 
the Bondi mass $M$ at null infinity using Weyl tensor $C_{abcd}$ as $M\sim\int 
rC_{urur}dS$ with $dS=r^{3}\sin\theta\cos\theta d\theta d\phi d\psi$. 
Since Eq.~(\ref{guu}) implies that $C_{urur}\sim V_{1}/r^{7/2}$ near null infinity,  
$M$ seems to diverge. 
Such a singular behavior of the Bondi mass has been pointed out in Ref.~\cite{Ishibashi:2007kb}. 
However, the solution of Eq.~(\ref{vsol}) implies that
the leading part of the integral $\int rC_{urur}dS$ vanishes.
That is to say, the finiteness of the Bondi mass is shown by solving the Einstein equations explicitly.
Using the solutions of Eqs.~(\ref{U11})-(\ref{U31}) and (\ref{vsol}), 
indeed, we can show the finiteness of the Bondi momentum.

\subsection{The Bondi angular momentum}

Next, let us define the Bondi angular momentum from $uA$ components of the metric.
Near null infinity, $g_{u\phi}$ and $g_{u\psi}$ are expanded as 
%
\begin{gather}
g_{u\phi} \,=\,\frac{1}{\sqrt{r}}\sin^{2}\theta U_{1}^{\phi} +\frac{1}{r}\sin^{2}\theta U_{2}^{\phi}
+\frac{1}{r\sqrt{r}}\sin^{2}\theta U_{3}^{\phi}+\frac{1}{r^2} j^{\phi} +O(r^{-5/2}) \\ 
g_{u\psi} \,=\,\frac{1}{\sqrt{r}}\cos^{2}\theta U_{1}^{\psi} +\frac{1}{r}\cos^{2}\theta U_{2}^{\psi} 
+\frac{1}{r\sqrt{r}}\cos^{2}\theta U_{3}^{\psi}+\frac{1}{r^2} j^{\psi} +O(r^{-5/2}), 
\end{gather} 
%
where 
%
\begin{gather}
j^{\phi}\,=\,\sin\theta D_{11}U_{1}^{\theta} + \sin^{2}\theta C_{21}U_{1}^{\phi} 
+\sin\theta\cos\theta D_{31}U_{1}^{\psi} +\sin^{2}\theta U_{4}^{\phi} \\
j^{\psi}\,=\,\cos\theta D_{21}U_{1}^{\theta} + \sin\theta\cos\theta D_{31}U_{1}^{\phi} 
-\cos^{2}\theta (C_{11}+C_{21})U_{1}^{\psi} +\cos^{2}\theta U_{4}^{\psi} .
\end{gather} 
%
Since $U_{1}^{\phi}$, $U_{1}^{\psi}$, $U_{2}^{\phi}$, $U_{2}^{\psi}$, 
$U_{3}^{\psi}$ and 
$U^{\phi}_{3}$ contain only total derivative terms on $S^3$~(see Eqs.~(\ref{U11})$-$(\ref{U33})), 
they cannot contribute to the definition of the global quantities. Therefore 
we define the Bondi angular momenta, $J_{\text{Bondi}}^{\phi}$ and $J_{\text{Bondi}}^{\psi}$, will be 
naturally defined by 
%
\begin{gather}
J_{\text{Bondi}}^{\phi}(u)\,=\,-\frac{1}{4\pi}\int_{S^{3}}j^{\phi} d\Omega \\
J_{\text{Bondi}}^{\psi}(u)\,=\,-\frac{1}{4\pi}\int_{S^{3}}j^{\psi} d\Omega .
\end{gather} 
%
From $R_{u\phi}\,=\,0$, we can derive the evolution equation for the angular momentum $J_{\text{Bondi}}^{\phi}(u)$ 
by gravitational waves as 
%
\begin{equation}
\frac{d}{du}J_{\text{Bondi}}^{\phi}(u)\,=\,-\frac{1}{4\pi}\int_{S^{3}}
\left[ \left(\frac{\partial j^{\phi}}{\partial u}\right)_{\text{radiation}} 
+\left(\frac{\partial j^{\phi}}{\partial u}\right)_{\text{total derivative}}
\right] d\Omega,
\end{equation} 
%
where $(\partial j^{\phi}/\partial u)_{\text{radiation}}$ is the radiation part given by
%
\begin{eqnarray}
\left(\frac{\partial j^{\phi}}{\partial u}\right)_{\text{radiation}}
&=&-\frac{1}{4}\frac{\partial C_{11}}{\partial\phi}
%
\frac{\partial C_{11}}{\partial u} -\frac{1}{8}\frac{\partial C_{21}}{\partial\phi}
\frac{\partial C_{11}}{\partial u} -\frac{1}{8}\frac{\partial C_{11}}{\partial\phi}
\frac{\partial C_{21}}{\partial u} -\frac{1}{4}\frac{\partial C_{21}}{\partial\phi}
\frac{\partial C_{21}}{\partial u}-\frac{1}{4}\frac{\partial D_{21}}{\partial\phi}
\frac{\partial D_{21}}{\partial u} 
\notag \\
&&
+\frac{1}{10}\tan\theta\frac{\partial D_{31}}
{\partial u}\left( \frac{\partial C_{11}}{\partial\psi}+
\frac{\partial C_{21}}{\partial\psi}\right) 
+\frac{3}{20}\tan\theta\frac{\partial D_{31}}{\partial\psi}
\frac{\partial C_{11}}{\partial u} +\frac{2}{5}\tan\theta\frac{\partial D_{31}}{\partial\psi}
\frac{\partial C_{21}}{\partial u}
\notag \\
&&
-\frac{1}{4}\frac{\partial D_{11}}{\partial\phi}\frac{\partial D_{11}}{\partial u}
+\frac{3}{20}\tan\theta\frac{\partial D_{11}}{\partial u}\frac{\partial D_{21}}{\partial\psi} -
\frac{3}{20}\tan\theta\frac{\partial D_{11}}{\partial\psi}\frac{\partial D_{21}}{\partial u} -
\frac{1}{4}\tan\theta\frac{\partial C_{11}}{\partial\psi}\frac{\partial D_{31}}{\partial u} 
\notag \\
&&
-\frac{1}{10}\tan\theta\frac{\partial C_{21}}{\partial u}\frac{\partial D_{31}}{\partial\psi} 
-\frac{2}{5}\tan\theta\frac{\partial D_{31}}{\partial u}\frac{\partial C_{21}}{\partial\psi} -
\frac{1}{4}\frac{\partial D_{31}}{\partial u}\frac{\partial D_{31}}{\partial\phi} 
\notag \\
&&
+\frac{3}{20\cos^{2}\theta}\frac{\partial D_{31}}{\partial u}\frac{\partial}{\partial\theta}
(\sin\theta\cos^{2}\theta D_{21})-\frac{3}{20}\cos\theta\frac{\partial}{\partial\theta}(\tan\theta D_{31})
\frac{\partial D_{21}}{\partial u} 
\notag \\
&&
+\frac{3}{20\cos^{2}\theta}\frac{\partial D_{11}}{\partial u}
\frac{\partial}{\partial\theta}(\sin\theta\cos^{2}\theta C_{11}) 
-\frac{3}{20}\cos\theta\frac{\partial C_{11}}{\partial u}\frac{\partial}{\partial\theta}(\tan\theta D_{11}) 
\notag \\
&&
+\frac{3}{20\sin\theta\cos\theta}\frac{\partial C_{21}}{\partial u}\frac{\partial}{\partial\theta}
(\sin^{2}\theta\cos\theta D_{11}) 
-\frac{3}{20}\sin\theta \frac{\partial C_{21}}{\partial\theta}
\frac{\partial D_{11}}{\partial u}
-\frac{3}{20\cos\theta}\frac{\partial}{\partial u}(C_{21}D_{11})
\end{eqnarray} 
%
and  $(\partial j^{\phi}/\partial u)_{\text{total derivative}}$ is the total 
derivative part which has no contribution to the evolution for the angular momentum 
by gravitational waves. The explicit form is given by
%
\begin{eqnarray}
\left(\frac{\partial j^{\phi}}{\partial u}\right)_{\text{total derivative}}
&=& \frac{1}{4}\left[\right. 3\sin^{2}\theta U_{2}^{\phi} 
-\tan^{2}\theta \frac{\partial^{2}}{\partial\psi^{2}} U_{2}^{\phi}
-\cot\theta\frac{\partial}{\partial\phi}U_{2}^{\theta}-\tan\theta\frac{\partial}{\partial\phi}U_{2}^{\theta}
+\frac{\partial^{2}}{\partial\phi\partial\theta} U_{2}^{\theta}+\frac{\partial}{\partial \phi}m \notag\\
&&
+\frac{\partial^2}{\partial\phi\partial\psi}U_{2}^{\psi} -\frac{1}{\sin\theta\cos\theta}\frac{\partial}{\partial\theta}
 \left( \sin^{3}\theta\cos\theta \frac{\partial}{\partial\theta} U_{2}^{\phi} \right) 
+\frac{1}{\sin\theta\cos\theta}\frac{\partial}{\partial\theta} \left( \sin^{2}\theta\cos\theta \frac{\partial D_{14}}{\partial u} 
\right) 
\notag \\
&&
+\tan\theta\frac{\partial}{\partial\psi}\frac{\partial D_{34}}{\partial u} 
+\frac{\partial}{\partial\phi}\frac{\partial C_{24}}{\partial u} 
-\frac{\partial}{\partial\phi}\frac{\partial B_{1}}{\partial u}
-\frac{3}{5}\tan\theta\frac{\partial}{\partial \psi}\left( D_{31}\frac{\partial C_{11}}{\partial u}\right)
-\frac{8}{5}\tan\theta\frac{\partial}{\partial \psi}\left( D_{31}\frac{\partial C_{21}}{\partial u}\right)
\notag \\
&&
+\frac{3}{5}\frac{\partial}{\partial \phi}\left( C_{21}\frac{\partial C_{21}}{\partial u}\right)
-\tan\theta\frac{\partial}{\partial \psi}\left( D_{21}\frac{\partial D_{11}}{\partial u}\right)
-\frac{2}{5}\frac{\partial}{\partial \phi}\left( D_{11}\frac{\partial D_{11}}{\partial u}\right)
+\frac{3}{5}\tan\theta\frac{\partial}{\partial \psi}\left( D_{11}\frac{\partial D_{21}}{\partial u}\right)
\notag \\
&&
+\tan\theta\frac{\partial}{\partial \psi}\left( C_{11}\frac{\partial D_{31}}{\partial u}\right)
+\frac{8}{5}\tan\theta\frac{\partial}{\partial \psi}\left( C_{21}\frac{\partial D_{31}}{\partial u}\right)
-\frac{2}{5}\frac{\partial}{\partial \phi}\left( D_{31}\frac{\partial D_{31}}{\partial u}\right) 
\notag \\
&&
-\frac{1}{\sin\theta\cos\theta}\frac{\partial}{\partial\theta}
  \left( \sin^{2}\theta\cos\theta D_{21}\frac{\partial D_{31}}{\partial u}  \right)
+\frac{3}{5\sin\theta\cos\theta}\frac{\partial}{\partial\theta}
  \left( \sin^{2}\theta\cos\theta D_{31}\frac{\partial D_{21}}{\partial u}  \right)
\notag \\
&&
-\frac{1}{\sin\theta\cos\theta}\frac{\partial}{\partial\theta}
  \left( \sin^{2}\theta\cos\theta D_{11}\frac{\partial C_{21}}{\partial u}  \right)
+\frac{3}{5\sin\theta\cos\theta}\frac{\partial}{\partial\theta}
  \left( \sin^{2}\theta\cos\theta D_{11}\frac{\partial C_{11}}{\partial u}  \right)
\notag \\
&&
-\frac{1}{\sin\theta\cos\theta}\frac{\partial}{\partial\theta}
  \left( \sin^{2}\theta\cos\theta C_{11}\frac{\partial D_{11}}{\partial u}  \right)
+\frac{3}{5\sin\theta\cos\theta}\frac{\partial}{\partial\theta}
  \left( \sin^{2}\theta\cos\theta C_{21}\frac{\partial D_{11}}{\partial u}  \right)
\left.\right].
\end{eqnarray} 
%
Then the evolution equation for angular momentum $J_{\text{Bondi}}^{\phi}(u)$ 
is expressed as
%
\begin{equation}
\frac{d}{du}J_{\text{Bondi}}^{\phi}(u)\,=\,-\frac{1}{4\pi}\int_{S^{3}}
\left(\frac{\partial j^{\phi}}{\partial u}\right)_{\text{radiation}} 
d\Omega . 
\end{equation} 
%
$(\partial j^{\phi}/\partial u)_{\text{total derivative}}$ plays an important role 
when showing the Poincare covariance of angular momentum $J^{\phi}_{\text{Bondi}}$ 
in the next section.
Note that
$dJ^{\phi}_{\text{Bondi}}/du=0$
if there is no gravitational wave, i.e., $\partial h^{(1)}_{AB}/\partial u =0$.

Similarly, from $R_{u\psi}=0$, we can see
%
\begin{equation}
\frac{d}{du}J_{\text{Bondi}}^{\psi}(u)\,=\,-\frac{1}{4\pi}\int_{S^{3}}
\left[\left(\frac{\partial j^{\psi}}{\partial u}\right) _{\text{radiation}} 
+\left( \frac{\partial j^{\psi}}{\partial u}\right) _{\text{total derivative}}
\right] d\Omega,
\end{equation} 
%
where
%
\begin{eqnarray}
\left(\frac{\partial j^{\psi}}{\partial u}\right) _{\text{radiation}}&=&
-\frac{1}{4}\frac{\partial C_{11}}{\partial\psi}\frac{\partial C_{11}}{\partial u}
-\frac{1}{4}\frac{\partial C_{21}}{\partial\psi}\frac{\partial C_{21}}{\partial u}
-\frac{1}{8}\frac{\partial C_{21}}{\partial\psi}\frac{\partial C_{11}}{\partial u}
-\frac{1}{8}\frac{\partial C_{11}}{\partial\psi}\frac{\partial C_{21}}{\partial u}
-\frac{3}{20}\cos\theta\frac{\partial D_{21}}{\partial\theta}\frac{\partial C_{11}}{\partial u}
\notag \\
&&
-\frac{1}{4}\cos\theta\frac{\partial C_{11}}{\partial\theta}\frac{\partial D_{21}}{\partial u}
-\frac{1}{4}\cot\theta\frac{\partial C_{21}}{\partial\phi}\frac{\partial D_{31}}{\partial u}
+\frac{3}{20}\cot\theta\frac{\partial C_{21}}{\partial\phi}\frac{\partial D_{31}}{\partial u}
-\frac{3}{20}\cot\theta\frac{\partial D_{31}}{\partial\phi}\frac{\partial C_{21}}{\partial u}
\notag \\
&&
-\frac{1}{4}\frac{\partial D_{11}}{\partial\psi}\frac{\partial D_{11}}{\partial u} 
+\frac{3}{20}\cot\theta\frac{\partial D_{11}}{\partial\phi}\frac{\partial D_{21}}{\partial u}
-\frac{3}{20}\cot\theta\frac{\partial D_{21}}{\partial\phi}\frac{\partial D_{11}}{\partial u}
-\frac{1}{4}\frac{\partial D_{21}}{\partial\psi}\frac{\partial D_{21}}{\partial u} 
-\frac{1}{4}\frac{\partial D_{31}}{\partial\psi}\frac{\partial D_{31}}{\partial u}
\notag \\
&&
-\frac{3}{20}\cot\theta\frac{\partial D_{31}}{\partial\phi}\frac{\partial}{\partial u}(C_{11}+C_{21})
+\frac{3}{20}\cot\theta\frac{\partial C_{11}}{\partial\phi}\frac{\partial D_{31}}{\partial u} 
+\frac{2}{5}\cot\theta\frac{\partial C_{21}}{\partial\phi}\frac{\partial D_{31}}{\partial u}
\notag \\
&&
+\frac{3}{20\sin\theta\cos\theta}\frac{\partial}{\partial\theta}
(\sin\theta\cos^{2}\theta C_{11})\frac{\partial D_{21}}{\partial u}
+\frac{3}{20\sin^{2}\theta}\frac{\partial}{\partial\theta}
(\sin^{2}\theta\cos\theta D_{11})\frac{\partial D_{31}}{\partial u}
\notag \\
&&
-\frac{3}{20}\sin\theta\frac{\partial}{\partial\theta}(\cot\theta D_{31})\frac{\partial D_{11}}{\partial u}
+\frac{2}{5}\cos\theta\frac{\partial C_{11}}{\partial\theta}\frac{\partial D_{21}}{\partial u}
+\frac{3}{20}\cos\theta\frac{\partial C_{21}}{\partial\theta}\frac{\partial D_{21}}{\partial u}
\notag \\
&&
-\frac{3}{20\sin\theta}\frac{\partial}{\partial u}(D_{21}C_{21})
-\frac{3}{20\sin\theta\cos\theta}\frac{\partial}{\partial\theta}(\sin\theta\cos^{2}\theta D_{21})
\frac{\partial}{\partial u}(C_{11}+C_{21})
,\end{eqnarray} 
%
and
%
\begin{eqnarray}
\left( \frac{\partial j^{\psi}}{\partial u}\right) _{\text{total derivative}}&=&
\frac{3}{4}\cos^{2}\theta U_{2}^{\psi} +\frac{\csc 2\theta}{2}\frac{\partial}{\partial \psi}U_{2}^{\theta}
+\frac{1}{4}\frac{\partial^2}{\partial\phi\partial\psi}U_{2}^{\phi} -\frac{\cot^{2}\theta}{4}
\frac{\partial^{2}}{\partial\phi^{2}}U^{\psi}_{2}+\frac{1}{4}\frac{\partial^{2}}{\partial\theta
\partial\psi}U^{\theta}_{2}+\frac{1}{4}\frac{\6 m}{\6\psi}
\notag \\
&&
-\frac{1}{4\sin\theta\cos\theta}\frac{\partial}{\partial\theta}
\left(\sin\theta\cos^{3}\theta\frac{\partial}{\partial\theta}U_{2}^{\psi}\right)
+\frac{1}{4\sin\theta\cos\theta}\left(\frac{\partial}{\partial\theta}\sin\theta\cos^{2}\theta 
\frac{\partial D_{24}}{\partial u}\right) +\frac{\cot\theta}{4}\frac{\partial^{2}}
{\partial\phi\partial u}D_{34}
\notag \\
&&
-\frac{1}{4}\frac{\partial^{2}}{\partial\psi\partial u}(C_{14}+C_{24}) 
-\frac{1}{4}\frac{\partial^{2}}{\partial\psi\partial u}B_{1} 
+\frac{3}{20}\frac{\partial}{\partial\psi}\left((C_{11}+C_{21})\frac{\partial}
{\partial u}(C_{11}+C_{21})\right)
\notag \\
&&
+\frac{1}{4}\frac{\partial}{\partial\phi}\left(\cot\theta D_{31}\frac{\partial}{\partial u}C_{11}\right)
+\frac{2}{5}\frac{\partial}{\partial\phi}\left(\cot\theta D_{31}\frac{\partial}{\partial u}C_{21}\right)
+\frac{1}{2}\frac{\partial}{\partial\psi}\left( D_{11}\frac{\partial}{\partial u}D_{11}\right)
\notag \\
&&
+\frac{3}{20}\frac{\partial}{\partial\phi}\left(\cot\theta D_{21}\frac{\partial}{\partial u}D_{11}\right)
+\frac{2}{5}\frac{\partial}{\partial\psi}\left( D_{21}\frac{\partial}{\partial u}D_{21}\right)
-\frac{1}{4}\frac{\partial}{\partial\phi}\left(\cot\theta D_{11}\frac{\partial}{\partial u}D_{21}\right)
\notag \\
&&
+\frac{2}{5}\frac{\partial}{\partial\psi}\left( D_{31}\frac{\partial}{\partial u}D_{31}\right)
-\frac{3}{20}\frac{\partial}{\partial\phi}\left(\cot\theta C_{11}\frac{\partial}{\partial u}D_{31}\right)
-\frac{2}{5}\frac{\partial}{\partial\phi}\left(\cot\theta C_{21}\frac{\partial}{\partial u}D_{31}\right)
\notag \\
&&
+\frac{2}{5\sin\theta\cos\theta}\frac{\partial}{\partial\theta}
\left(\sin\theta\cos^{2}\theta D_{21}\frac{\partial}{\partial u}C_{11}\right)
+\frac{1}{4\sin\theta\cos\theta}\frac{\partial}{\partial\theta}
\left(\sin\theta\cos^{2}\theta D_{21}\frac{\partial}{\partial u}C_{21}\right)
\notag \\
&&
+\frac{3}{20\sin\theta\cos\theta}\frac{\partial}{\partial\theta}
\left(\sin\theta\cos^{2}\theta D_{31}\frac{\partial}{\partial u}D_{11}\right)
-\frac{2}{5\sin\theta\cos\theta}\frac{\partial}{\partial\theta}
\left(\sin\theta\cos^{2}\theta C_{11}\frac{\partial}{\partial u}D_{21}\right)
\notag \\
&&
-\frac{3}{20\sin\theta\cos\theta}\frac{\partial}{\partial\theta}
\left(\sin\theta\cos^{2}\theta C_{21}\frac{\partial}{\partial u}D_{21}\right)
-\frac{1}{4\sin\theta\cos\theta}\frac{\partial}{\partial\theta}
\left(\sin\theta\cos^{2}\theta D_{11}\frac{\partial}{\partial u}D_{31}\right).
\end{eqnarray} 
%
Then the evolution equation for the angular momentum $J_{\text{Bondi}}^{\psi}(u)$  
is given by
%
\begin{equation}
\frac{d}{du}J_{\text{Bondi}}^{\psi}(u)\,=\,-\frac{1}{4\pi}\int_{S^{3}}
\left(\frac{\partial j^{\psi}}{\partial u}\right)_{\text{radiation}} 
d\Omega . 
\end{equation} 
%

\section{Asymptotic symmetry}
\label{Sec:AS}

In this section, we consider asymptotic symmetry at null infinity. 
Following our previous work \cite{Tanabe:2009va}, we 
first discuss the asymptotic symmetry. See Ref.~\cite{Tanabe:2009va} 
for the details. Then we will show the Poincare covariance of the Bondi mass 
and angular momentum. 

\subsection{Preliminary}

Asymptotic symmetry is defined as the transformation group which 
preserve the boundary conditions of Eqs.~(\ref{bbc1})-(\ref{bbc2})
at null infinity. 
By infinitesimal transformations $\xi^a$, the metric is 
transformed as $g_{ab}\rightarrow g_{ab} +\delta g_{ab}$, where
%
\begin{equation}
\delta g_{ab}\,=\,\nabla_{a}\xi_{b} +\nabla_{b}\xi_{a}.
\end{equation} 
%
To preserve the boundary conditions, the metric variation $\delta g_{ab}$ should satisfy
following conditions
%
\begin{gather}
\delta g_{rr}\,=\,0\,,\,\delta g_{rA}\,=\,0\,,\,g^{AB}\delta g_{AB}\,=\,0\label{bc1} \\
\delta g_{uu}\,=\,O(r^{-3/2})\,,\,\delta g_{uA}\,=\,O(r^{-1/2})\,,\,
\delta g_{AB}\,=\,O(r^{1/2})\label{bc2}. 
\end{gather} 
%
The conditions of Eq.~(\ref{bc1}) comes from the definition of the Bondi coordinates. 
Next the conditions of Eq.~(\ref{bc2}) are required from the behavior of gravitational fields near 
null infinity. From the condition of Eq.~(\ref{bc1}), we can see that 
the infinitesimal transformation $\xi^a$ can be written  as
%
\begin{eqnarray}
&&\xi_{r} \,=\, f(u,x^{A})e^{B} \\
&&\xi_{B}g^{AB} \,=\, f^{A}(u,x^{A}) -f(u,x^{A})U^{A}+\int^{\infty}_{r}dr' e^{B}
\frac{\partial f}{\partial x^{B}}g^{AB} \\
&&\xi_{u} \,=\, -\frac{re^{B}}{3}\left( -\frac{\partial\xi_{A}}{\partial x^{B}} 
+\xi_{C}\Gamma^{C}_{AB}+\xi_{r}\Gamma^{r}_{AB}\right)g^{AB},
\end{eqnarray} 
%
where $f(u,x^{A})$ and $f^{A}(u,x^{A})$ are functions which satisfy
%
\begin{eqnarray}
&& \frac{\partial f^{A}}{\partial u}\,=\,0 \label{confkill1}\\
&& \mathcal{D}_{A}f_{B} +\mathcal{D}_{B}f_{A}\,=\,-2\frac{\partial f}{\partial u}h^{(0)}_{AB} 
\label{confkill2}\\
&& \mathcal{D}_{A}\mathcal{D}_{B}f\,=\,\frac{1}{3}\mathcal{D}^{2}f h^{(0)}_{AB}
\label{feq}. 
\end{eqnarray} 
%
In the above, $\mathcal{D}_{A}$ is the covariant derivative with respect to 
$h^{(0)}_{AB}$. 
From Eqs.~(\ref{confkill1}) and (\ref{confkill2}), we can see that 
$f^{A}$ should be conformal Killing 
vector on $S^{3}$. Since such transformation group (conformal Killing group 
on $S^{3}$) is isomorphic to the Lorentz group, $f^{A}$ corresponds to 
the generator of the Lorentz 
transformations. The function $f$ can be written by $f=-F(x^{A})u/3+\alpha(x^{A})$,
where $F(x^{A})=\mathcal{D}_{A}f^{A}$. We would guess 
that $\alpha(x^{A})$ is the generator 
of translations. From Eq.~(\ref{feq}), we can see that 
the function $\alpha(x^{A})$ can be written as
%
\begin{equation}
\alpha(x^{A})\,=\,a_{u} + a_{x}\sin\theta\cos\phi +a_{y}\sin\theta\sin\phi
+a_{z}\cos\theta\cos\psi +a_{w}\cos\theta\sin\psi.
\end{equation} 
%
This $\alpha$ has only five parameters which correspond to five directions of 
translations. Thus, the asymptotic symmetry generated by $\xi$ at null infinity 
is the Poincare group which is semi-direct of the Lorentz group and the translation 
group. On the other hand, in four dimensions, there is no conditions on $f$ 
like Eq.~(\ref{feq}). This means that $\alpha(x^{A})$ in $f$ generates the infinite dimensional 
translation group, that is, the supertranslation group in four dimensions.   

In our previous work \cite{Tanabe:2009va}, 
we have not discussed the asymptotic symmetry 
in terms of global charges. Hence we will address this point in the next subsection. 

\subsection{Poincare covariance of the Bondi mass}

Asymptotic quantities like the Bondi mass and angular momentum should be 
global charges associated to the asymptotic symmetry.
To confirm this, in this section, 
we will consider translations generated by $\xi$ with 
%
\begin{equation}
f\,=\,\alpha(x^{A})\,,\,f^{A}\,=\,0 \label{trans}.
\end{equation} 
%
This transformation implies $u\rightarrow u-f(x^{A})$.

The global charges associated with asymptotic symmetry (the Poincare group) are 
energy-momentum vector $P_{a}$ and angular momentum $M_{ab}$. And they 
should be transformed under translations of Eq.~(\ref{trans}) as
%
\begin{equation}
P_{a}\rightarrow P_{a}
\,,\,
\qquad
M_{ab}\rightarrow M_{ab}+2P_{[a}\omega_{b]},
\end{equation} 
%
where $f=\hat{x}^{a}\omega_{a}$ and $\hat{x}^{a}=(1,\hat{x}^{i})$. However, 
we are considering dynamical space-times which has no exact timelike Killing 
vector. This means that 
the quantities $P_{a}$ and $M_{ab}$ would change due to gravitational waves 
under translations ($u\rightarrow u-f(x^{A})$). Then, the expected transformations 
of $P_{a}$ and $M_{ab}$ under translation of Eq.~(\ref{trans}) are
%
\begin{eqnarray}
P_{a}(u)&\rightarrow & P_{a}(u-f)\,=\,
                      P_{a}(u) -\left(f\frac{d}{du}P_{a}(u)\right)_{\text{radiation}} \label{mass}\\
M_{ab}(u)&\rightarrow &M_{ab}(u-f)+2P_{[a}(u)\omega_{b]}\,=\,
                         M_{ab}(u)+2P_{[a}(u)\omega_{b]} -\left(f\frac{d}{du}M_{ab}(u)\right)
_{\text{radiation}} \label{angular}.
\end{eqnarray} 
%
In Eq.~(\ref{angular}), note that the argument of the second term 
$2P_{[a}\omega_{b]}$ is $u$, not $u-f$. This is because 
this term corresponds to the orbital angular momentum 
generated by translations. The purpose of this section is to confirm this 
property. First, we check the relation of Eq.~(\ref{mass}).  
Under the translations of Eq.~(\ref{trans}), $m(u,x^{A})$ in $g_{uu}$ 
transforms as
%
\begin{eqnarray}
m\rightarrow m-\alpha(x^{A})\frac{\partial m}{\partial u}
-\frac{2}{3}\frac{\partial \alpha}{\partial\theta}\frac{\partial U^{\theta}_{2}}{\partial u}
-\frac{2}{3}\frac{\partial \alpha}{\partial\phi}\frac{\partial U^{\phi}_{2}}{\partial u}
-\frac{2}{3}\frac{\partial \alpha}{\partial\psi}\frac{\partial U^{\psi}_{2}}{\partial u}
+(\text{total derivative terms}).
\end{eqnarray} 
%
Here we do not write down the explicit form of the total derivative 
terms in this equation because it is slightly complicated 
and not important for our purpose. From the above equation, 
we find that the Bondi energy momentum $P^{a}_{\text{Bondi}}$ transforms under the translations as
%
\begin{eqnarray}
P^{a}_{\text{Bondi}}&\rightarrow & P^{a}_{\text{Bondi}}+\frac{3}{16\pi}\int_{S^{3}}
\hat{x}^{a}\left[-\alpha(x^{A})\frac{\partial m}{\partial u}
-\frac{2}{3}\frac{\partial \alpha}{\partial\theta}\frac{\partial U^{\theta}_{2}}{\partial u}
-\frac{2}{3}\frac{\partial \alpha}{\partial\phi}\frac{\partial U^{\phi}_{2}}{\partial u}
-\frac{2}{3}\frac{\partial \alpha}{\partial\psi}\frac{\partial U^{\psi}_{2}}{\partial u}
 \right]d\Omega
\notag \\
&=&P^{a}_{\text{Bondi}}+\frac{1}{16\pi}\int_{S^{3}}\hat{x}^{a}\left\{\alpha\left[ 
\left(\frac{\partial C_{11}}{\partial u}\right)^2 +\frac{\partial C_{11}}
{\partial u}\frac{\partial C_{21}}{\partial u} +\left(\frac{\partial C_{21}}{\partial u}\right)^2 
\right.\right.
\notag \\
&&~~~~~~~~~~~~~~~~~~~~~~~~~~~~~\left.
~
+\left(\frac{\partial D_{11}}{\partial u}\right)^2 
+\left(\frac{\partial D_{21}}{\partial u}\right)^2 +\left(\frac{\partial D_{31}}
{\partial u}\right)^2 \right] \notag\\
&&~~~~~~~~~~~~~~~~~~~\left. 
\qquad 
-\left[
\frac{2}{\sin\theta\cos\theta}\frac{\partial}{\partial\theta}\left(\alpha\sin\theta\cos\theta 
\frac{\partial U^{\theta}_{2}}{\partial u}\right) 
+2\frac{\partial^{2}}{\partial\phi\partial u}(\alpha U^{\phi}_{2})+2\frac{\partial^{2}}
{\partial\psi\partial u}(\alpha U^{\psi}_{2})\right]\right\}d\Omega
\notag\\
&=&P^{a}_{\text{Bondi}}+\frac{1}{16\pi}\int_{S^{3}}\alpha\hat{x}^{a}\left\{ 
\left(\frac{\partial C_{11}}{\partial u}\right)^2 +\frac{\partial C_{11}}
{\partial u}\frac{\partial C_{21}}{\partial u} +\left(\frac{\partial C_{21}}{\partial u}\right)^2 
\right.\notag \\
&&~~~~~~~~~~~~~~~~~~~~~~~~~~~~~~~~\left.
+\left(\frac{\partial D_{11}}{\partial u}\right)^2 
+\left(\frac{\partial D_{21}}{\partial u}\right)^2 +\left(\frac{\partial D_{31}}
{\partial u}\right)^2 \right\}d\Omega. \label{masstrans}
\end{eqnarray} 
%
In the above 
we used the fact that $U^{A}_{2}$ are time independent. This can be shown using 
Eqs.~(\ref{U12})-(\ref{U32}) and $\partial h^{(2)}_{AB}/\partial u=0$. 
As seen in section~\ref{Sec:AQ}, the integral part in Eq.~(\ref{masstrans}) can be interpreted as
the energy momentum loss by gravitational wave $(fdP_{a}/du)_{\text{radiation}}$. Then, the Bondi energy 
momentum
satisfy the Poincare covariance of Eq.~(\ref{mass}).

\subsection{Poincare covariance of the Bondi angular momentum}

Next, we show the Poincare covariance of the Bondi angular momentum $J_{\text{Bondi}}^{\phi}$. 
By the translations of Eq~(\ref{trans}), $j^{\phi}$ is transformed as
%
\begin{eqnarray}
j^{\phi}&\rightarrow& j^{\phi}-\alpha(x^{A})\frac{\partial j^{\phi}}{\partial u}
+(\delta j^{\phi})_{\text{non radiation}}
+(\text{total derivative terms}),
\end{eqnarray} 
%
where 
%
\begin{eqnarray}
(\delta j^{\phi})_{\text{non radiation}}&=&
\frac{2}{3}\frac{\partial^2\alpha}{\partial\theta\partial\phi}U^{\theta}_{2}
-\frac{1}{3}\frac{\partial\alpha}{\partial\theta}\frac{\partial U^{\theta}_{2}}{\partial\phi}
+\frac{1}{3}\tan^{2}\theta\frac{\partial^{2}\alpha}{\partial\psi^{2}}U^{\phi}_{2}
+\frac{\partial^{2}\alpha}{\partial\phi^{2}}U^{\phi}_{2}
+\frac{1}{6}(6+8\cos 2\theta)\tan\theta\frac{\partial\alpha}{\partial\theta}U^{\phi}_{2}
\notag \\
&&
+\frac{1}{3}\sin^{2}\theta\frac{\partial^{2}\alpha}{\partial\theta^{2}}U^{\phi}_{2}
+\tan^{2}\theta\frac{\partial\alpha}{\partial\psi}\frac{\partial U^{\phi}_{2}}{\partial\psi}
+\frac{2}{3}\frac{\partial\alpha}{\partial\phi}\frac{\partial U^{\phi}_{2}}{\partial\phi}
+\sin^{2}\theta\frac{\partial\alpha}{\partial\theta}\frac{\partial U^{\phi}_{2}}{\partial\theta}
+\frac{2}{3}\frac{\partial^{2}\alpha}{\partial\phi\partial\psi}U^{\psi}_{2}
-\frac{1}{3}\frac{\partial\alpha}{\partial\psi}\frac{\partial U^{\psi}_{2}}{\partial\phi}
\notag \\
&&
-m\frac{\partial \alpha}{\partial\phi}
+\frac{1}{4}\frac{\partial\alpha}{\partial\phi}\frac{\partial B_{1}}{\partial u}
-\frac{1}{4}\frac{\partial\alpha}{\partial\phi}\frac{\partial C_{24}}{\partial u}
-\frac{1}{4}\frac{\partial\alpha}{\partial\theta}\frac{\partial D_{14}}{\partial u}
-\frac{1}{4}\tan\theta\frac{\partial\alpha}{\partial\psi}\frac{\partial D_{34}}{\partial u}
+\frac{3}{20}\tan\theta D_{31}\frac{\partial\alpha}{\partial\psi}\frac{\partial C_{11}}{\partial u}
\notag \\
&&
-\frac{3}{20}\sin\theta D_{11}\frac{\partial\alpha}{\partial\theta}\frac{\partial C_{11}}{\partial u}
+\frac{2}{5}\tan\theta D_{31}\frac{\partial\alpha}{\partial\psi}\frac{\partial C_{21}}{\partial u}
-\frac{3}{20} C_{21}\frac{\partial\alpha}{\partial\phi}\frac{\partial C_{21}}{\partial u}
+\frac{1}{4}\sin\theta D_{11}\frac{\partial\alpha}{\partial\theta}\frac{\partial C_{21}}{\partial u}
\notag \\
&&
+\frac{1}{4}\tan\theta D_{21}\frac{\partial\alpha}{\partial\psi}\frac{\partial D_{11}}{\partial u}
+\frac{1}{10}D_{11}\frac{\partial\alpha}{\partial\phi}\frac{\partial D_{11}}{\partial u}
+\frac{1}{4}\sin\theta C_{11}\frac{\partial\alpha}{\partial\theta}\frac{\partial D_{11}}{\partial u}
-\frac{3}{20}\sin\theta C_{21}\frac{\partial\alpha}{\partial\theta}\frac{\partial D_{11}}{\partial u}
\notag \\
&&
-\frac{3}{20}\tan\theta D_{11}\frac{\partial\alpha}{\partial\psi}\frac{\partial D_{21}}{\partial u}
-\frac{3}{20}\sin\theta D_{31}\frac{\partial\alpha}{\partial\theta}\frac{\partial D_{21}}{\partial u}
-\frac{1}{4}\tan\theta C_{11}\frac{\partial\alpha}{\partial\psi}\frac{\partial D_{31}}{\partial u}
-\frac{2}{5}\tan\theta C_{21}\frac{\partial\alpha}{\partial\psi}\frac{\partial D_{31}}{\partial u}
\notag \\
&&
+\frac{1}{10}D_{31}\frac{\partial\alpha}{\partial\phi}\frac{\partial D_{31}}{\partial u}
+\frac{1}{4}\sin\theta D_{21}\frac{\partial\alpha}{\partial\theta}\frac{\partial D_{31}}{\partial u}.
\end{eqnarray} 
%
Then, under the translations of Eq.~(\ref{trans}), 
the Bondi angular momentum at null infinity transforms as
%
\begin{eqnarray}
J^{\phi}_{\text{Bondi}}&\rightarrow& J^{\phi}_{\text{Bondi}} 
-\frac{1}{4\pi}\int_{S^{3}}\left[-\alpha(x^{A})\frac{\partial j^{\phi}}{\partial u}
+(\delta j^{\phi})_{\text{non radiation}}\right]d\Omega
\notag \\
&&
=J^{\phi}_{\text{Bondi}}
-\frac{1}{4\pi}\int_{S^{3}}\left[-\alpha(x^{A})\left(\frac{\partial j^{\phi}}{\partial u}
\right)_{\text{radiation}}-\alpha(x^{A})\left(\frac{\partial j^{\phi}}{\partial u}
\right)_{\text{total derivative}} 
+(\delta j^{\phi})_{\text{non radiation}}\right]d\Omega.
\end{eqnarray} 
%
Now, the straightforward calculations tells us  
%
\begin{eqnarray}
-\alpha(x^{A})\left(\frac{\partial j^{\phi}}{\partial u}\right)_{\text{total derivative}}
+(\delta j^{\phi})_{\text{non radiation}}=-\frac{3}{4}m\frac{\partial\alpha}{\partial\phi}
+(\text{supermomentum})_{\phi} +(\text{total derivative}),
\end{eqnarray} 
%
where supermomentum term is given by
%
\begin{eqnarray}
(\text{supermomentum})_{\phi}&=&
\frac{\alpha}{4}\left[5\cot\theta\frac{\partial}{\partial\phi}U^{\theta}_{2}
-3\tan\theta\frac{\partial}{\partial\phi}U^{\theta}_{2}
+3\frac{\partial^{2}}{\partial\theta\6\phi}U^{\theta}_{2}\right] \notag \\
&&
+\frac{\alpha}{12}\left[23\sin^{2}\theta U^{\phi}_{2} -32\cos^{2}\theta U^{\phi}_{2}
-31\cos\theta\sin\theta\frac{\partial}{\partial\theta}U^{\phi}_{2}
+5\frac{\sin^{3}\theta}{\cos\theta}\frac{\partial}{\partial\theta}U^{\phi}_{2}
\right.\notag \\
&&~~~~~~~~~~~~~~~~~~~~~~~~~~~~~~~~~~~\left.
-5\sin^{2}\theta\frac{\partial^{2}}{\partial\theta^{2}}U^{\phi}_{2}
+4\frac{\partial^{2}}{\partial\phi^{2}}U^{\phi}_{2}
-5\tan^{2}\theta\frac{\partial^{2}}{\partial\psi^{2}}U^{\phi}_{2}\right]
\notag \\
&&
+\frac{3}{4}\alpha\frac{\partial^{2}}{\partial\phi\partial\psi}U^{\psi}_{2}
.
\end{eqnarray} 
%
In the above $\alpha(x^{A})$ has only $l=0$ and $l=1$ modes of scalar harmonics on $S^{3}$. 
Using the solution of the Einstein equations~(\ref{U12})$-$(\ref{U32}), then, 
we can show that these supermomentum terms become total derivative.
Then, taking these
results altogether, we see 
%
\begin{eqnarray}
J^{\phi}_{\text{Bondi}}&\rightarrow& J^{\phi}_{\text{Bondi}} 
-\frac{1}{4\pi}\int_{S^{3}}\left[-\alpha(x^{A})\left(\frac{\partial j^{\phi}}{\partial u}
\right)_{\text{radiation}}\right]d\Omega
+\frac{3}{16\pi}\int_{S^{3}}m(u,x^{A})\frac{\partial\alpha}{\partial\phi}
d\Omega.
\end{eqnarray} 
%
This transformation is equivalent to $M_{\hat{x}\hat{y}}\rightarrow M_{\hat{x}\hat{y}}
+2P_{[\hat{x}}\omega_{\hat{y}]}
-(f dM_{\hat{x}\hat{y}}/du)_{\text{radiation}}$. This stands for 
the Poincare covariance of the Bondi angular momentum. 
In the same way, we can show the Poincare covariance of 
angular momentum $J_{\text{Bondi}}^{\psi}$. Under translations, $J_{\text{Bondi}}^{\psi}$
transforms as
%
\begin{eqnarray}
J^{\psi}_{\text{Bondi}}&\rightarrow& J^{\psi}_{\text{Bondi}} 
-\frac{1}{4\pi}\int_{S^{3}}\left[-\alpha(x^{A})\left(\frac{\partial j^{\psi}}{\partial u}
\right)_{\text{radiation}}+(\text{supermomentum})_{\psi}\right]d\Omega
+\frac{3}{16\pi}\int_{S^{3}}m(u,x^{A})\frac{\partial\alpha}{\partial\psi}
d\Omega,
\end{eqnarray} 
%
where the supermomentum term is
%
\begin{eqnarray}
(\text{supermomentum})_{\psi}&=&
\frac{\alpha}{4}\left[3\cot\theta\frac{\partial}{\partial\psi}U^{\theta}_{2}
-5\tan\theta\frac{\partial}{\partial\psi}U^{\theta}_{2}
+3\frac{\partial^{2}}{\partial\theta\psi}U^{\theta}_{2}\right] 
-\frac{3}{4}\alpha\frac{\partial^{2}}{\partial\phi\partial\psi}U^{\phi}_{2}
\notag \\
&&
+\frac{\alpha}{12}\left[-32\sin^{2}\theta U^{\psi}_{2} +23\cos^{2}\theta U^{\psi}_{2}
+31\cos\theta\sin\theta\frac{\partial}{\partial\theta}U^{\psi}_{2}
-5\frac{\cos^{3}\theta}{\sin\theta}\frac{\partial}{\partial\theta}U^{\psi}_{2}
\right.\notag \\
&&~~~~~~~~~~~~~~~~~~~~~~~~~~~~~~~~~~~\left.
-5\cos^{2}\theta\frac{\partial^{2}}{\partial\theta^{2}}U^{\psi}_{2}
+4\frac{\partial^{2}}{\partial\psi^{2}}U^{\psi}_{2}
-5\cot^{2}\theta\frac{\partial^{2}}{\partial\phi^{2}}U^{\phi}_{2}\right].
\end{eqnarray} 
%
For the $l=0$ and $l=1$ modes in $\alpha$, these supermomentum terms 
become total derivative. Finally we obtain the Poincare transformations 
of angular momentum as 
%
\begin{eqnarray}
J^{\psi}_{\text{Bondi}}&\rightarrow& J^{\psi}_{\text{Bondi}} 
-\frac{1}{4\pi}\int_{S^{3}}\left[-\alpha(x^{A})\left(\frac{\partial j^{\psi}}{\partial u}
\right)_{\text{radiation}}\right]d\Omega
+\frac{3}{16\pi}\int_{S^{3}}m(u,x^{A})\frac{\partial\alpha}{\partial\psi}
d\Omega.
\end{eqnarray} 
%
This transformation is equivalent to $M_{\hat{z}\hat{w}}\rightarrow M_{\hat{z}\hat{w}}+
2P_{[\hat{z}}\omega_{\hat{w}]}-(f dM_{\hat{z}\hat{w}}/du)_{\text{radiation}}$.  

Since $\alpha$ contains all $l$ mode of scalar harmonics on $S^{2}$ in four dimensions, 
the supermomentum does not vanish in general. Thus we cannot show the Poincare 
covariance of angular momentum in four dimensions~(See Appendix A for the details).

\section{summary and discussion}
\label{Sec:Summary}

In this paper, we defined the Bondi angular momentum at null infinity in five 
dimensions and showed the Poincare covariance of the Bondi mass and angular momentum. 
In addition, we successfully confirmed 
the Bondi mass loss and angular momentum loss/gain due to gravitational wave. 

Asymptotic symmetry at null infinity is an infinite dimensional 
translational group (supertranslations) in four dimensions, not a four 
dimensional group. Then this implies that the angular 
momentum at null infinity has always ambiguities. 
Contrasted with this,  it is shown that asymptotic symmetry at 
null infinity is the Poincare group in five dimensions. Then we can define the Bondi angular 
momentum at null infinity in a Poincare covariant way 
without any ambiguities. 

There are remaining issues. 
In this paper we focused on the five dimensional space-times. We would expect that 
our approach can be extended to higher dimensions than five. However, there is a 
critical point, that is, we had to introduce the concrete angular coordinate to 
solve the Einstein equation. On the other hand, one does not want to use the 
concrete angular coordinates when one is interested in higher dimensions. We have to 
resolve this troublesome issue. It is also interesting to study 
the asymptotic structure at null infinity in even dimensions using the Bondi coordinate 
because the finiteness has not been shown in even dimensions. They are left for future works.

\section*{Acknowledgment}
KT is supported by JSPS Grant-Aid for Scientific Research (No.~21-2105). 
N.T. is
supported by the DOE Grant DE-FG03-91ER40674, and thank Takahiro Tanaka for his grateful aid.
T.S. is partially supported by Grant-Aid for Scientific Research from Ministry of Education, Science,
Sports and Culture of Japan (Grant Nos. 21244033, 21111006, 20540258, and 19GS0219). 
This work is also supported by the Grant-
in-Aid for the Global COE Program ``The Next Generation of Physics, Spun from Universality 
and Emergence'' from the Ministry of Education, Culture, Sports, Science and Technology (MEXT) 
of Japan.

\appendix

\section{Angular momentum at null infinity in four dimensions}

In this Appendix we discuss the angular momentum at null infinity in four dimensions using 
the Bondi coordinates. This will be useful for the comparison with five dimensional cases. 
We could not find old studies on angular momentum based on the Bondi coordinate. 
Here we will show that the angular momentum has always supertranslational ambiguities, which is
represented by supermomentum. 

\subsection{Bondi coordinate and Einstein equations in four dimensions}

In this section we introduce the Bondi coordinates and solve the Einstein equations 
in four dimensions \cite{Bondi:1962px,Sachs:1962wk}. In the Bondi coordinates
$(u,r,\theta,\phi)$ the metric can be written as
%
\begin{eqnarray}
ds^{2}\,=\,-\frac{V{e^{B}}}{r}du^{2}-2e^{B}dudr+r^{2}h_{AB}(dx^{A}+U^{A}du)(dx^{B}+U^{B}du),
\end{eqnarray} 
%
where
%
\begin{equation}
h_{AB}\,=\,
\begin{pmatrix}
e^{C} & \sin\theta\sinh D \\
\sin\theta\sinh D & e^{E}\sin^{2}\theta 
\end{pmatrix}.
\end{equation} 
%
There is a gauge condition such that $\det h_{AB} =\sin^{2}\theta$, and then $e^{E}$ can be written 
by $C$ and $D$ as
%
\begin{equation}
e^{E}\,=\,e^{-C}(1+\sinh^{2}D).
\end{equation} 
%
As in five dimensions, the functions $C$ and $D$ represent the degree of 
freedom of gravitational field. 
Given the function $C$ and $D$ on $u=\text{const.}$ hypersurfaces, the other metric functions $B$, $U^{A}$
and $V$ are determined by the Einstein equations. 
To solve the Einstein equations near null infinity, we expand the function $C$ and $D$ 
as
%
\begin{eqnarray}
C(u,r,x^{A})\,=\,\frac{C_{1}(u,x^{A})}{r}+\frac{C_{2}(u,x^{A})}{r^{2}}+O(r^{-3}) \\
D(u,r,x^{A})\,=\,\frac{D_{1}(u,x^{A})}{r}+\frac{D_{2}(u,x^{A})}{r^{2}}+O(r^{-3}) .
\end{eqnarray} 
%
Then, using the Einstein equations $R_{ab}=0$, the metric functions $V$, $B$ and $U^{A}$ can be 
written by $C$ and $D$. From $R_{rr}=0$, 
%
\begin{eqnarray}
B(u,r,x^{A})\,=\,\frac{B_{1}(u,x^{A})}{r^2}+O(r^{-3}) \\
B_{1}(u,x^{A})\,=\,-\frac{1}{8}(C_{1}^{2}+D_{1}^{2}).
\end{eqnarray} 
%
From $R_{rA}=0$, 
%
\begin{eqnarray}
U^{A}(u,x^{A})\,=\,\frac{U_{1}^{A}(u,x^{A})}{r^2}+\frac{U_{2}^{A}(u,x^{A})}{r^3} +O(r^{-4}) \\
U_{1}^{\theta}\,=\,\frac{1}{2\sin^{2}\theta}\left(\frac{\partial}{\partial\theta}(\sin^{2}\theta C_{1})
                  +\frac{\partial}{\partial\phi}(\sin\theta D_{1})\right) \\
\sin^{2}\theta U_{1}^{\phi}\,=\,\frac{1}{2}\left(\frac{1}{\sin\theta}\frac{\partial}{\partial\theta}(\sin^{2}\theta D_{1})
                  -\frac{\partial}{\partial\phi}C_{1}\right) .
\end{eqnarray} 
%
From $h^{AB}R_{AB}=0$, 
%
\begin{eqnarray}
\frac{V}{r}
\,=\,1-\frac{m(u,x^{A})}{r}+O(r^{-2}).
\end{eqnarray} 
%
Since the functions $U_{2}^{A}(u,x^{A})$ and $m(u,x^{A})$ are the integration constants 
in the $r$-integration, they are free functions of $(u,x^{A})$.

\subsection{Bondi mass and angular momentum}

Now, $g_{uu}$ is expanded as 
%
\begin{eqnarray}
g_{uu}=-1+\frac{m(u,x^{A})}{r}+O(r^{-2}),
\end{eqnarray} 
%
and then we define the Bondi mass $M_{\text{Bondi}}$ 
and the Bondi momentum $P^i_\text{Bondi}$ in four dimensions as%
\footnote{The coefficients of the definitions are determined so that 
these quantities are coincide to ADM quantities at spatial infinity. }
%
\begin{eqnarray}
M_{\text{Bondi}}(u)\,=\,\frac{1}{8\pi}\int_{S^{2}}m(u,x^{A})d\Omega ,
\\
P^{i}_{\text{Bondi}}(u)\,=\,\frac{1}{8\pi}\int_{S^{2}}m(u,x^{A})\hat{x}^{i}d\Omega ,
\end{eqnarray} 
%
where $d\Omega=\sin\theta d\theta d\phi$ and $\hat{x}^{i}=
(\hat{x},\hat{y},\hat{z})=(\sin\theta\cos\phi,\sin\theta\sin\phi,\cos\theta)$. 

From the Einstein equation $R_{uu}=0$, we can obtain the Bondi mass loss law 
by gravitational waves as 
%
\begin{eqnarray}
\frac{d}{du}M_{\text{Bondi}}(u)&=&-\frac{1}{16\pi}\int_{S^{2}}\left[
\left(\frac{\partial C_{1}}{\partial u}\right)^{2} +\left(\frac{\partial D_{1}}{\partial u}\right)^{2} 
-\frac{1}{\sin\theta}\frac{\partial}{\partial\theta}\left(\sin\theta \frac{\partial U^{\theta}_{1}}
{\partial u}\right)-\frac{\partial^{2}}
{\partial\phi\partial u}
U^{\phi}_{1}
\right] d\Omega\notag\\
&<&0.
\end{eqnarray} 
%
Thus, the Bondi mass in four dimensions is always decreased by gravitational waves. 

$g_{u\phi}$ are expanded as 
%
\begin{eqnarray}
g_{u\phi}\,=\,
\sin^2\theta
U_{1}^{\phi} +\frac{\sin\theta D_{1}U_{1}^{\theta} -\sin^{2}\theta C_{1}U_{1}^{\phi}
+\sin^{2}\theta U_{2}^{\phi}}{r}+O(r^{-2}),
\end{eqnarray} 
%
and then we define the Bondi angular momentum in four dimensions as
%
\begin{eqnarray}
J_{\text{Bondi}}(u)\,=\,-\frac{3}{16\pi}\int_{S^{2}}(\sin\theta D_{1}U_{1}^{\theta} -\sin^{2}\theta C_{1}U_{1}^{\phi}
                                +\sin^{2}\theta U_{2}^{\phi})d\Omega .
\end{eqnarray} 
%
From the Einstein equation $R_{u\phi}=0$, we 
can see that the evolution equation for the Bondi angular momentum becomes
%
\begin{eqnarray}
\frac{d}{du}J_{\text{Bondi}}\,=\,-\frac{3}{16\pi}\int_{S^{2}}\left[
\left(\frac{\partial j}{\partial u}\right)_{\text{radiation}}+\left(\frac{\partial j}{\partial u}\right)_{\text{total derivative}}\right]
d\Omega,
\end{eqnarray} 
%
where $(\partial j/\partial u)_{\text{radiation}}$ is the radiation part given by
%
\begin{eqnarray}
\left(\frac{\partial j}{\partial u}\right)_{\text{radiation}}&=&
-\frac{1}{6\sin\theta}\frac{\partial C_{1}}{\partial u}\frac{\partial}{\partial\theta}(\sin^{2}\theta D_{1})
+\frac{1}{6\sin\theta}\frac{\partial D_{1}}{\partial u}\frac{\partial}{\partial\theta}(\sin^{2}\theta C_{1})
-\frac{1}{3}\frac{\partial D_{1}}{\partial u}\frac{\partial D_{1}}{\partial\phi}
-
\frac{1}{3}\frac{\partial C_{1}}{\partial u}\frac{\partial C_{1}}{\partial\phi}
\notag \\
&&
-\frac{\sin\theta}{6}\frac{\partial C_{1}}{\partial u}\frac{\partial D_{1}}{\partial\theta}
+\frac{\sin\theta}{6}\frac{\partial D_{1}}{\partial u}\frac{\partial C_{1}}{\partial\theta}
,
\end{eqnarray} 
%
and $(\partial j/\partial u)_{\text{total derivative}}$ is total derivative given by
%
\begin{eqnarray}
\left(\frac{\partial j}{\partial u}\right)_{\text{total derivative}}&=&
\frac{1}{3}
\frac{\partial}{\partial\phi}m-\frac{\cot\theta}{3}\frac{\partial}{\partial\phi}U_{1}^{\theta}
+\frac{1}{3}\frac{\partial^{2}}{\partial\phi\partial\theta}U^{\theta}_{1} 
+\frac{2\sin^{2}\theta}{3}U_{1}^{\phi}
-\frac{1}{3\sin\theta}\frac{\6}{\6\theta}\left(
\sin^3\theta\frac{\6}{\6\theta}U^\phi_1
\right)
-\frac{1}{3}\frac{\partial^{2}}{\partial u\partial\phi}C_{2}
\notag \\
&&
+\frac{1}{3\sin\theta}\frac{\partial}{\partial\theta}\left(\sin^{2}\theta\frac{\partial D_{2}}{\partial u}\right)
-\frac{1}{3}\frac{\partial^{2}}{\partial u\partial\phi}B_{1}
+\frac{1}{2}\frac{\partial}{\partial\phi}\left(D_{1}\frac{\partial}{\partial u}D_{1}\right)
+\frac{1}{6}\frac{\partial}{\partial\phi}\left(C_{1}\frac{\partial}{\partial u}C_{1}\right)
\notag\\
&&
-\frac{1}{2\sin\theta}\frac{\partial}{\partial\theta}\left(\sin^{2}\theta C_{1}\frac{\partial}{\partial u}D_{1}\right)
+\frac{1}{2\sin\theta}\frac{\partial}{\partial\theta}\left(\sin^{2}\theta D_{1}\frac{\partial}{\partial u}C_{1}\right).
\end{eqnarray} 
%
Thus, $(\partial j/\partial u)_{\text{total derivative}}$ has no contribution to the angular momentum 
loss by gravitational waves.

\subsection{Asymptotic symmetry and supermomentum}

Asymptotic symmetry is transformation group which satisfy the conditions given by
%
\begin{gather}
\delta g_{rr}\,=\,0\,,\,\delta g_{rA}\,=\,0\,,\,\,g^{AB}\delta g_{AB}\,=\,0 \label{BC14} \\
\delta g_{uu}\,=\,O(r^{-1})\,,\,\delta g_{uA}\,=\,O(1)\,,\,\delta g_{AB}\,=\,O(r) \label{BC24},
\end{gather} 
%
where $\delta g_{ab}=\nabla_{a}\xi_{b}+\nabla_{b}\xi_{a}$ is the infinitesimal transformation by $\xi$. 
These are required to keep the asymptotic behavior of the metric to be unchanged.

From the conditions (\ref{BC14}), which come from the definition of the Bondi coordinate, 
we can check that the components of $\xi^a$ should 
have following form
%
\begin{eqnarray}
&&\xi_{r} \,=\, f(u,x^{A})e^{B} \\
&&\xi_{B}g^{AB} \,=\, f^{A}(u,x^{A}) -f(u,x^{A})U^{A}+\int^{\infty}_{r}dr' e^{B}
\frac{\partial f}{\partial x^{B}}g^{AB} \\
&&\xi_{u} \,=\, -\frac{re^{B}}{2}\left( -\frac{\partial\xi_{A}}{\partial x^{B}} 
+\xi_{C}\Gamma^{C}_{AB}+\xi_{r}\Gamma^{r}_{AB}\right)g^{AB}.
\end{eqnarray} 
%
From the conditions (\ref{BC24}), which are required from the behavior of the gravitational fields near 
null infinity, $f^{A}$ should satisfy
%
\begin{gather}
\frac{\partial}{\partial u}f^{A}\,=\,0 \\
\mathcal{D}_{A}f_{B}+\mathcal{D}_{B}f_{A}\,=\,-2\frac{\partial f}{\partial u}h^{(0)}_{AB} \label{A27},
\end{gather} 
%
where $\mathcal{D}_{A}$ is the covariant derivative with $h^{(0)}_{AB}$ given by
%
\begin{equation}
h^{(0)}_{AB}\,=\,
\begin{pmatrix}
1&0 \\
0&\sin^{2}\theta
\end{pmatrix}.
\end{equation} 
%
$f^{A}$ generate the conformal transformation group on $S^{2}$ and such group is isomorphic to 
the Lorentz group in four dimensions. Then $f^{A}$ stands for the generator of the Lorentz transformations. 
Contrasted with in five dimensions, there are no further conditions on $f$ in four dimensions. 
Thus Eq.~(\ref{A27}) tells us that $f$ can be written as $f=-(u/2)D^{A}f_{A}+\alpha(x^{A})$. 
$\alpha(x^{A})$ is arbitrary 
function on $S^{2}$ which is called supertranslations. 

Now, we consider the transformation of the Bondi mass $M_{\text{Bondi}}$ by supertranslations 
$f=\alpha(x^{A})$. $M_{\text{Bondi}}$ is transformed as
%
\begin{eqnarray}
M_{\text{Bondi}}&\rightarrow& M_{\text{Bondi}} +\frac{1}{8\pi}\int_{S^{2}}
\left[-f\frac{\partial m}{\partial u} -\frac{\partial f}{\partial\theta}\frac{\partial U_{1}^{\theta}}{\partial u}
-\frac{\partial f}{\partial\phi}\frac{\partial U_{1}^{\phi}}{\partial u}   \right]
\notag \\
&&
=M_{\text{Bondi}}+\frac{1}{16\pi}\int_{S^{2}}f\left[
\left(\frac{\partial C_{1}}{\partial u}\right)^{2} +\left(\frac{\partial D_{1}}{\partial u}\right)^{2} 
\right].
\end{eqnarray} 
%
This is the Poincare transformation under the presence of gravitational waves. 
In four dimensions we cannot show the Poincare covariance of the Bondi momentum because 
$\partial U_{1}^{A}/\partial u\neq 0$ in the presence of gravitational waves.
This means that the Bondi momentum has supertranslational ambiguities in four 
dimensions.

Next, we consider the transformations of the Bondi angular momentum by supertranslations. 
Angular momentum $J_{\text{Bondi}}$ is transformed as
%
\begin{eqnarray}
J_{\text{Bondi}}&\rightarrow & J_{\text{Bondi}} -\frac{3}{16\pi}\int_{S^{2}}\left[
-f\frac{\partial}{\partial u}(\sin\theta D_{1}U^{\theta}_{1}-\sin^{2}\theta C_{1}U^{\phi}_{1}+\sin^{2}\theta U^{\phi}_{2})
-m\frac{\partial f}{\partial\phi}+\frac{1}{2}U^{\theta}_{1}\frac{\partial^{2}}{\partial\theta\partial\phi}f
-\frac{1}{2}\frac{\partial f}{\partial \theta}\frac{\partial}{\partial\phi}U^{\theta}_{1}
\right.\notag \\
&&~~~~~~~~~~~~~~~~~~~~~~~~~~~~~~~~
+\frac{1}{2}
\frac{\6 f}{\6 \phi}
\frac{\partial}{\partial\phi}U^{\phi}_{1}
+
\sin^2\theta
\frac{\partial f}{\partial\theta}\frac{\partial}{\partial\theta}U^{\phi}_{1}
+\frac{1}{2}U^{\phi}_{1}
\frac{\partial^{2}f}{\partial\phi^{2}}
+2\sin\theta\cos\theta U^{\phi}_{1}\frac{\partial f}{\partial\theta}
+\frac{1}{3}
\frac{\6 C_2}{\6u}
\frac{\partial f}{\partial\phi}
\notag \\
&&~~~~~~~~~~~~~~~~~~~~~~~~~~~~~~~~
-\frac{1}{3}\sin\theta 
\frac{\6 D_2}{\6 u}\frac{\partial f}{\partial\theta}
+\frac{1}{3}
\frac{\6 B_1}{\6 u}
\frac{\partial f}{\partial\phi}
-\frac{1}{2}\frac{\partial f}{\partial\phi}D_{1}\frac{\partial D_{1}}{\partial u}
-\frac{1}{6}\frac{\partial f}{\partial\phi}C_{1}\frac{\partial C_{1}}{\partial u}
+\frac{1}{2}\sin\theta
\frac{\partial f}{\partial\theta}
C_{1}\frac{\partial D_{1}}{\partial u}
\notag \\
&&~~~~~~~~~~~~~~~~~~~~~~~~~~~~~~~~~~~~~~~~~~~~~~~~~~~~~~~~~~~~~\left.
-\frac{1}{2}\sin\theta
\frac{\partial f}{\partial\theta}
D_{1}\frac{\partial C_{1}}{\partial u}
+(\text{total derivative term})
\right] d\Omega
\notag \\
&=&
J_{\text{Bondi}} -\frac{3}{16\pi}\int_{S^{2}}\left[ 
-f\left(\frac{\partial j}{\partial u}\right)_{\text{radiation}}+ (\text{supermomentum}) \right]d\Omega
+\frac{1}{8\pi}\int_{S^{2}}m\frac{\partial f}{\partial\phi}d\Omega ,
\end{eqnarray} 
%
where supermomentum term is
%
\begin{eqnarray}
(\text{supermomentum})&=&
\frac{2}{3}f\left(2\cot\theta \frac{\partial}{\partial\phi}U^{\theta}_{1}+
\frac{\partial^{2}}{\partial\theta\partial\phi}U^{\theta}_{1}\right) 
\notag \\
&&~~~~~~~~~~~~~
-\frac{2}{3}f\left( 6\cos^{2}\theta U^{\phi}_{1} -2\sin^{2}\theta U^{\phi}_{1}
+6\sin\theta\cos\theta\frac{\partial}{\partial\theta}U^{\phi}_{1}
+\sin^{2}\theta\frac{\partial^{2}}{\partial\theta^{2}}U^{\phi}_{1}\right).
\end{eqnarray} 
%
If we take $f$ as $l=1$ mode of scalar harmonics on $S^{2}$, the supermomentum terms become to have 
the total derivative form
and then  
we can obtain the Poincare covariance of the angular momentum $M_{\hat{x}\hat{y}}
\rightarrow M_{\hat{x}\hat{y}}+2P_{[\hat{x}}\omega_{\hat{y}]}-f(dj/du)_{\text{radiation}}$. However, 
for $l>1$ mode, the supermomentum terms do not become total derivative form. This means that 
we cannot obtain the Poincare covariance of 
angular momentum. Under the presence of gravitational waves, we cannot restrict $f$ to $l=1$ mode. 
In general, therefore, the 
angular momentum has supertranslational ambiguity in four dimensions.



\end{document}